\documentclass[mathleft
]{an}
\usepackage{graphicx}
\usepackage{times}
\usepackage{natbib}
\usepackage{sidecap}
\begin{document}

\Pagespan{789}{}
\Yearpublication{2006}%
\Yearsubmission{2005}%
\Month{11}%
\Volume{999}%
\Issue{88}%

\newcommand{\sml}{{\sc SpotModeL}~}
\newcommand{\HA}{H$\alpha$~}

\title{Four-colour photometry of EY Dra: a study of an ultra-fast rotating active dM1-2e star}

\author{K. Vida\inst{1}, K. Ol\'ah\inst{1}, Zs. K\H{o}v\'ari\inst{1}, J. Jurcsik\inst{1}, \'A. S\'odor\inst{1}, M. V\'aradi\inst{1,4}, B. Belucz\inst{2}, I. D\'ek\'any\inst{1}, Zs. Hurta\inst{1},  I. Nagy\inst{2}, K. Posztob\'anyi\inst{3}}

\titlerunning{$BV(RI)_C$ photometry of EY Dra}
\authorrunning{K. Vida et al.}
\institute{
Konkoly Observatory of the Hungarian Academy of Sciences, H-1525 Budapest, PO Box 67, Hungary
\and 
E\"otv\"os Lor\'and University, Department of Astronomy, H-1518 Budapest, PO Box 32, Hungary
\and
AEKI, KFKI Atomic Energy Research Institute, Thermohydraulic Department, H–1525 Budapest 114, PO Box 49, Hungary
\and
Observatoire de Gen\`{e}ve, Universite de Gen\`{e}ve, CH–1290, Sauverny, Switzerland
}


\abstract{
{We present more than 1000-day long photometry of EY Draconis in $BV(RI)_C$ passbands. The changes in the light curve are caused by the spottedness of the rotating surface. Modelling of the spotted surface shows  that there are two large active regions present on the star on the opposite hemispheres. The evolution of the surface patterns suggests a flip-flop phenomenon.
Using Fourier analysis, we detect a rotation period of $P_\mathrm{rot}=0.45875$d, and an activity cycle with $P\approx350$d, similar to the 11-year long cycle of the Sun. This cycle with its year-long period is the shortest one ever detected on active stars. 
Two bright flares are also detected and analysed. }
}  
\maketitle
\sloppy
\section{Introduction}
Fast rotating late-type stars are known to possess strong magnetic fields  \citep[see e.g.][]{rot-mag}. These manifest in observable features like starspots, flares, emission in the H$\alpha$ and Ca {\sc ii} lines, activity cycles, and significant emission in the EUV/X-ray regime  \citep[see e.g.][]{review,review-ca,euv,rot-act}.
Although  Doppler-imaging and polarimetry gives a priceless tool in the hands of astronomers, the value of long-term photometric observations should not be underestimated, since this is the only way to continuously follow the behaviour of a star on the timescale of years or decades.

EY Dra is a well-known example of active stars with a history reaching back to almost two decades.
In 1991 the ROSAT EUV/X-ray satellite detected EY Dra as an EUV source. 
\cite{jeffries} classified the object as a single, rapidly rotating ($v\sin i=61 \mathrm{km s}^{-1}$) dM1--2e star, similar to HK Aqr. The authors presented a thorough spectroscopic analysis and recommended the star as a candidate for studies of extreme magnetic activity. 
\cite{eibe} carried out high-resolution \HA measurements, and  suggested the presence of plage-like regions and prominence clouds above the stellar surface.
\cite{barnes} used Doppler-imaging to obtain a surface map. The Doppler-map showed spots on all latitudes, but no significant evidence was found for the existence of a polar cap. Although active regions on the Sun appear always on lower latitudes, high-latitude spots and polar caps are familiar features on rapidly rotating stars as a result of the dominance of the Coriolis force over buoyancy force on the rising flux tubes \citep{polar_cap}.

The first extensive photometric study of EY Dra was done by \cite{robb}, who examined $V$-band data. The light curve showed an unusual W-shape, a small flare, and a period of 0.459d.
Recent study of \cite{korhonen} used both $V$ and $R$ photometric measurements and spectroscopic data in optical and infrared domain. The authors find two active regions, chromospheric plages and prominences. The plages seemed to be associated with the active regions, as seen on the Sun.
In \cite{eydra} $V$-band photometry was presented covering almost 500d. The data indicated slow evolution of the surface spots and also a possible longer cycle of about 300d. The length of the dataset, however, was too short for such a conclusive statement.
\cite{spitzer} analyzed 24 $\mu$m and 70 $\mu$m Multiband Imaging Photometer for Spitzer (MIPS) observations of 70 dwarfs, including EY Dra. They found no IR excess, and concluded there is no disk around the star.

In this paper we carry out the analysis of $BV(RI)_C$ photometry of EY Dra covering more than 1100 days.

\section{Observations}
\begin{figure}
\centering
\includegraphics[width=0.40\textwidth]{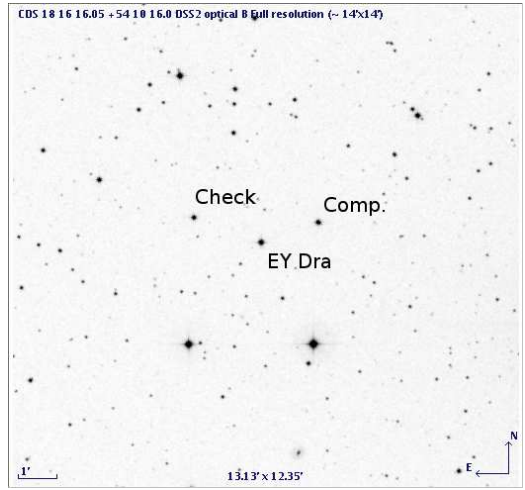}
\caption{Finding chart for the EY Dra field.}
\label{fig:chart}
\end{figure}

\begin{figure}
\centering
\includegraphics[width=0.47\textwidth]{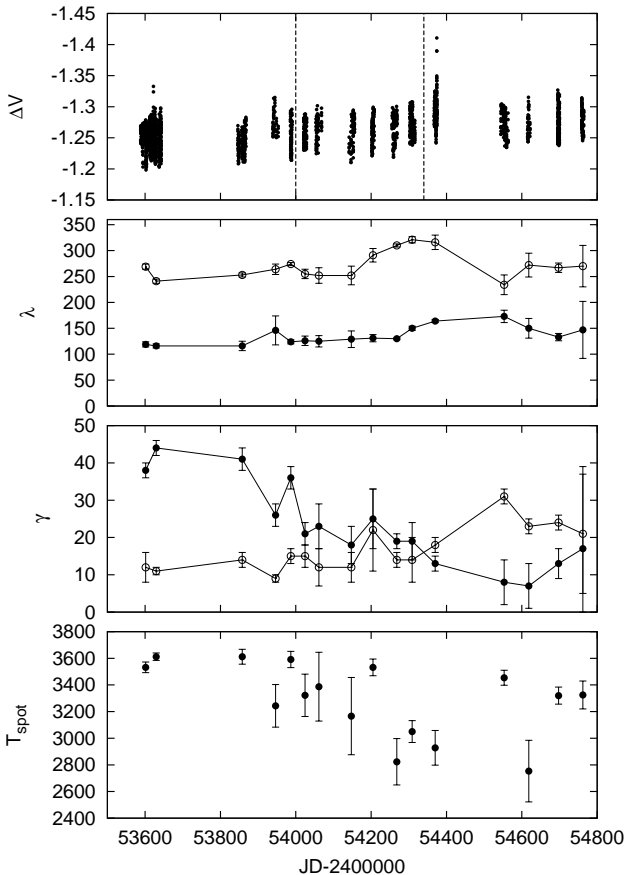}
\caption{From top to bottom:  Light curve of EY Draconis in $V$ passband. The brighter points over the light curve at JDs 2453622 and 2454374 are results of flares. Dashed line shows where the data was split for Fourier analysis, as described in Section \ref{sect:fourier}.  Results of the spot modelling, showing the spot longitudes, sizes ($\lambda$ and $\gamma$, respectively) and spot temperatures. }
\label{fig:allv}
\end{figure}

Observations were obtained using the 60cm telescope of the Konkoly Observatory at Sv\'abhegy, Budapest equipped with a Wright Instruments $750\times1100$ CCD camera (FoV $17'\times24'$). Measurements were carried out on 116 nights between 2005 August 5 and 2008 October 24 using $BV(RI)_C$ filters. Altogether more than 2500 data points were collected in each passband. Data reduction was carried out using standard IRAF
\footnote{IRAF is distributed by the National Optical Astronomy Observatory, which is operated by the Association of Universities for Research in Astronomy, Inc., under cooperative agreement with the National Science Foundation.}
packages. Differential aperture photometry was done using DAOPHOT package.

GSC 03904-00259 and GSC 03904-00645 were used as comparison and check star, respectively (see finding chart in Fig. \ref{fig:chart}).
Comparison-check magnitudes are 
$-0.304\pm0.018$,
$-0.316\pm0.014$,
$-0.347\pm0.020$ and
$-0.348\pm0.022$
for $B$, $V$, $R_C$, and $I_C$ filters, respectively.
Unfortunately, no standard magnitudes are available for the comparison star, so we used standardized differential magnitudes in the paper.

The resulting $V$ light curve is plotted in the top panel of Fig. \ref{fig:allv}.
For the phased light curves we used the ephemeris 
$E=2453588.16582+0.4587$d.

\section{Analysis}
\subsection{Period search}
\label{sect:fourier}
\begin{figure}
\centering
\includegraphics[angle=-90,width=0.48\textwidth]{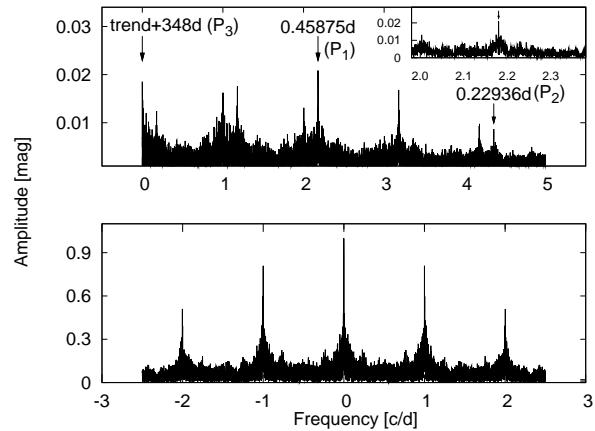}
\caption{Fourier spectrum of the EY Dra $V$ light curve (top) and the spectral window (bottom). The insert in the top plot shows the power spectrum zoomed in to $P_\mathrm{rot}$.}
\label{fig:mufran}
\end{figure}

\begin{figure}
\includegraphics[angle=-90,width=0.48\textwidth]{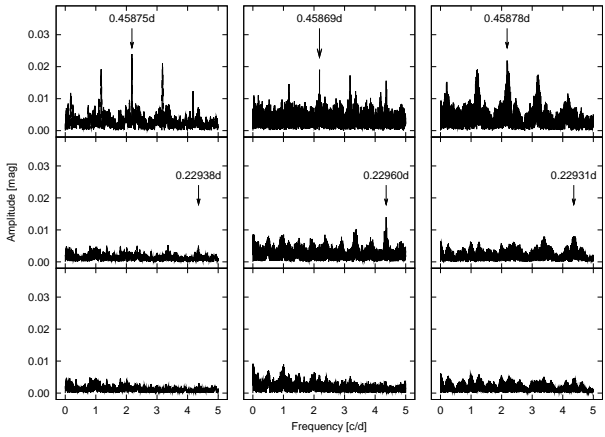}
\caption{Fourier analysis of the original $V$ light curve split to three parts. The second row shows the Fourier-spectrum after prewhitening with $P_\mathrm{rot}$, in the third row both $P_\mathrm{rot}$ and the signal at $P_\mathrm{rot}/2$ is removed. The same rotation period within the errors is found in all three parts. A signal at the half period is also present in all three segments. This suggests two major active regions on the stellar surface all the time.}
\label{fig:split}
\end{figure}

\begin{figure}
\includegraphics[angle=-90,width=0.47\textwidth]{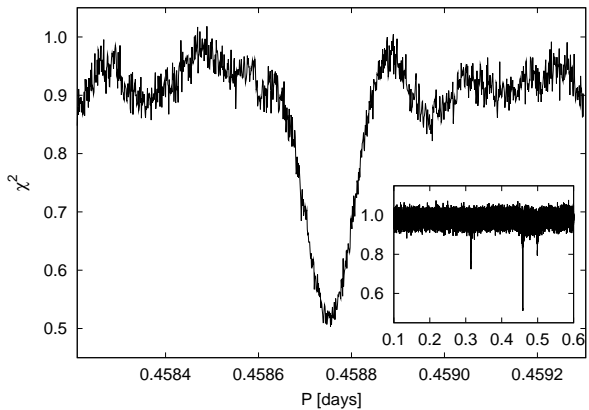}
\caption{Periodogram made by using the SLLK method described by \cite{sllk}. The large plot is zoomed in to $P_\mathrm{rot}=0.45875$d, the insert shows the full periodogram in the range of  0.1--0.6d.}
\label{fig:sllk}
\end{figure}

Fourier analysis of the data was performed using MUFRAN ({\sc MUlti FRe\-quency ANalysis}) written by \cite{mufran}. 
The Fourier spectrum and the spectral window for the $V$ light curve are plotted on Fig. \ref{fig:mufran}. 
The strongest signal is caused by the rotational modulation at $P_1=P_\mathrm{rot}=0.45875\pm0.00003$d.
The errors of the periods were estimated by increasing the residual scatter of
the least-squares solutions by 0 $.\!\!^{\rm m}$0005 which corresponds to 10\%
of the precision of the data used \citep[cf.][]{error}.
Another signal at the half of the rotation period ($P_2=0.22936\pm0.00001$d) is also present, and the result is the same when we fit the data with the main period and its first harmonic together. This result reflects two major active regions on the stellar surface in the opposite hemispheres.  (see the 'W'-shaped light curves on Fig. \ref{fig:slices} and the derived spot longitudes from the modelling in Fig. \ref{fig:allv})
After removing these signals plus a long-term trend which is comparable to the length of the dataset, we find a period of $P_3=348\pm19$d. Small residuals are present near the main period and its half, possibly originating from the modulation residuals, i.e., from the changes of the light curve shapes.

To check for changes in the periods, we divided the data to three parts, each covering $\approx400$ days, as shown in the top panel of Fig. \ref{fig:allv}, and performed Fourier analysis for each segment. The result is plotted in Fig. \ref{fig:split}. The resulting frequencies differ only a few seconds, and the differences are beyond the reliability of the period determination ($0.0001$d), so the period is stable during the observations.

As a test, we checked the data for periods using the SLLK method (String/Rope length method using Lafler--Kinman statistic) described by \cite{sllk}. This method phases the light curves with different periods, and selects the period giving the smoothest light curve as the correct one. This method is very useful for finding periods for stars having non-sinusoidal light curve shape. The SLLK method gave the same result as the Fourier analysis ($P_\mathrm{rot}=0.45875$d, see the periodogram in Fig. \ref{fig:sllk}).

\subsection{Spot modelling}

\label{sect:analysis}
\begin{figure*}
\centering
\includegraphics[angle=-90,width=.94\textwidth]{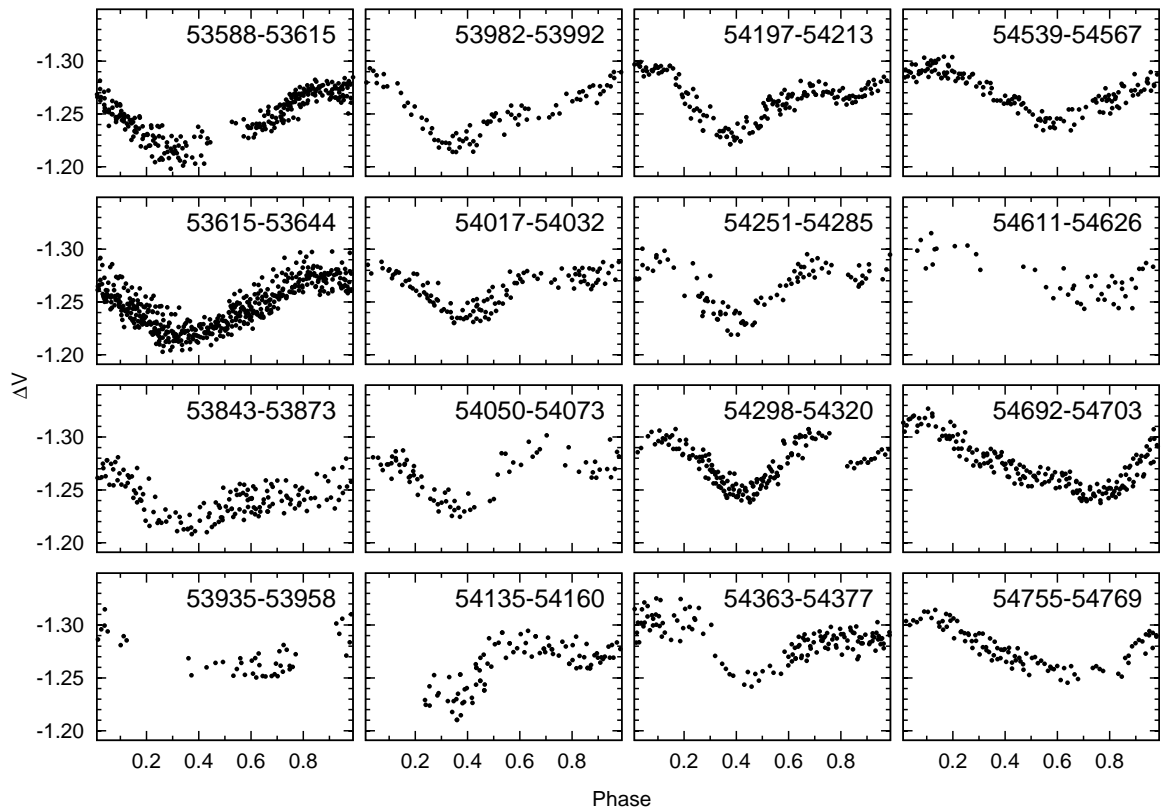}
\caption{Phased light curves of EY Draconis in $V$ passband. These light curves served as input for the spot modelling. }
\label{fig:slices}
\end{figure*}

\begin{figure*}
\centering
\includegraphics[angle=-90,width=.94\textwidth]{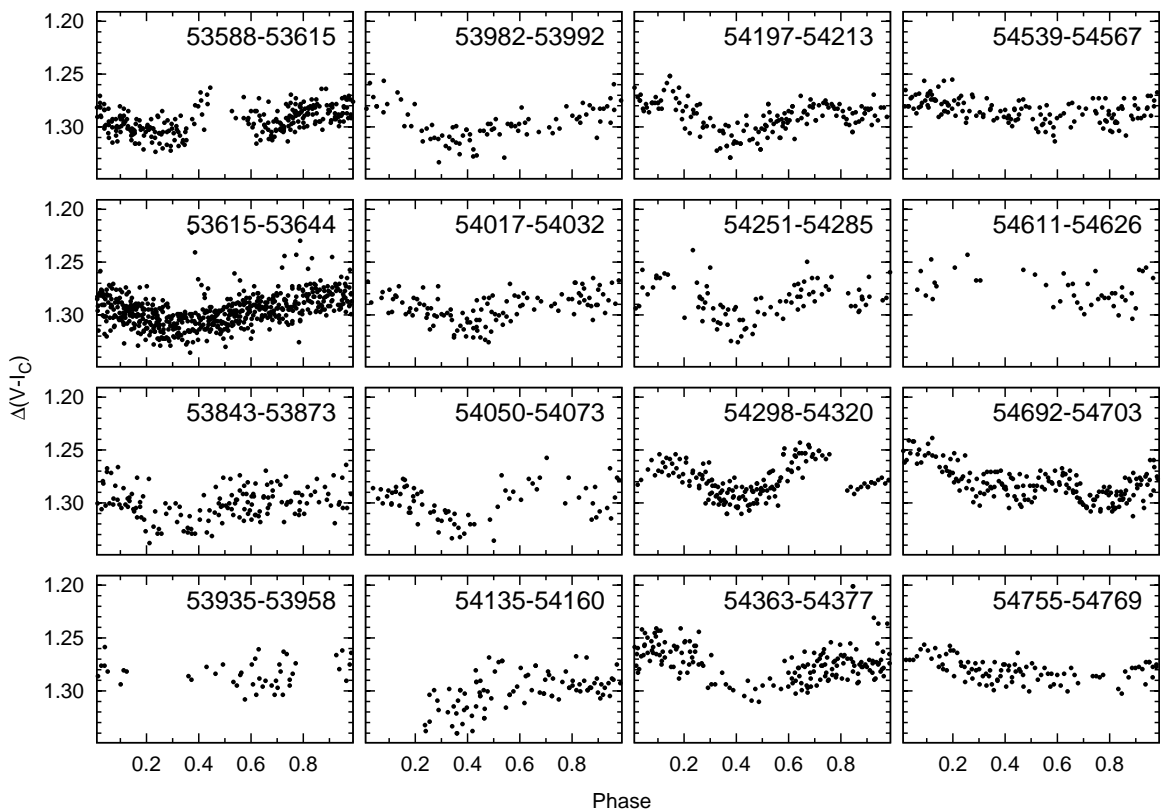}
\caption{Phased $V-I_C$ colour index curves of EY Draconis. These curves were used to model the spot temperatures.}
\label{fig:slices_ci}
\end{figure*}

\begin{figure}
\centering
\includegraphics[angle=0,width=.241\textwidth]{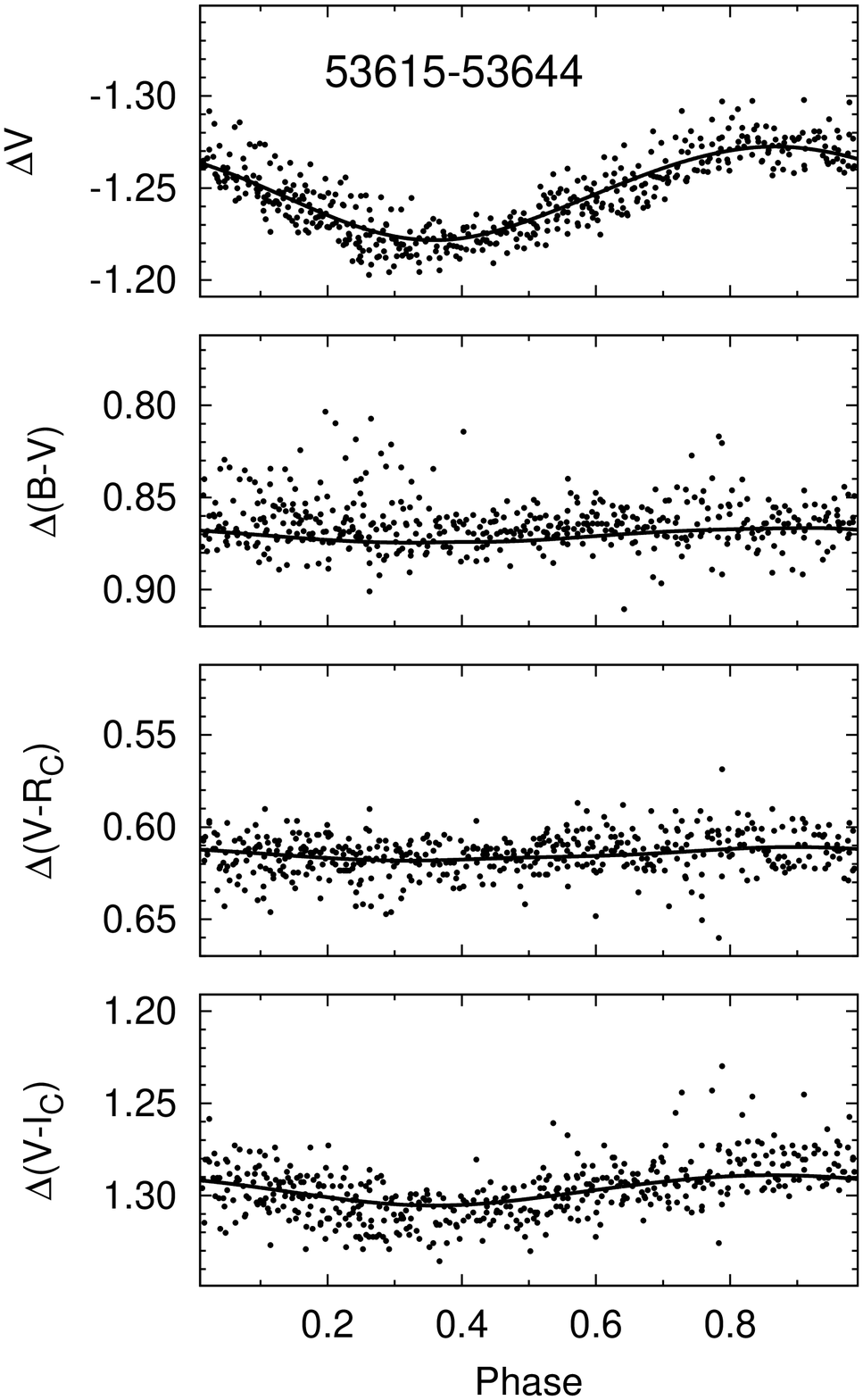}
\includegraphics[angle=0,width=.241\textwidth]{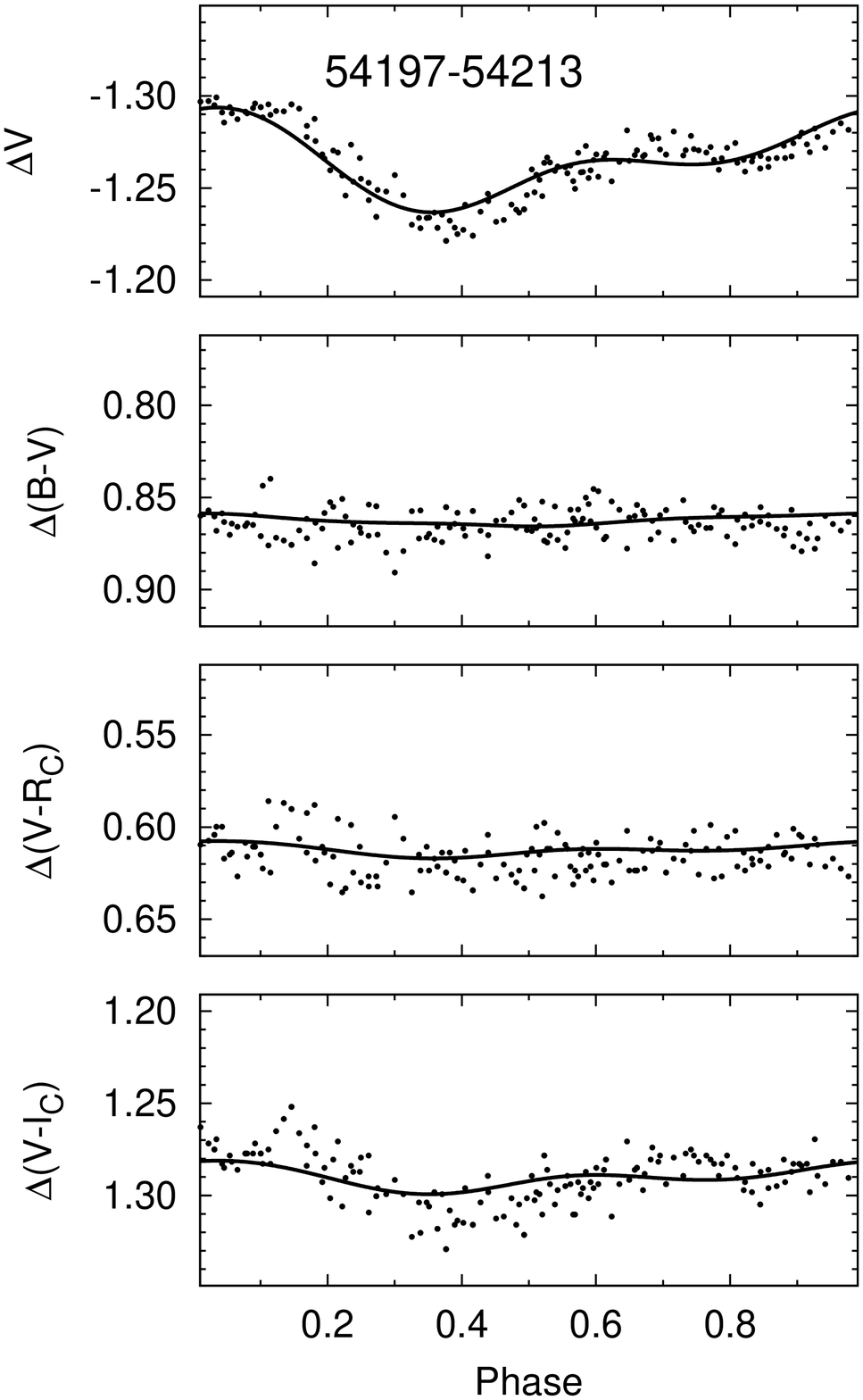}
\caption{Typical $V$ light curves and colour index curves of EY Draconis showing two states of the surface with one and two large active regions. Continuous line shows the fitted spot model for $V-I_C$ colour index. Note the tiny colour index amplitudes.}
\label{fig:smlfit}
\end{figure}
\begin{table}
\caption{Basic stellar parameters for EY Dra}
\begin{tabular}{ccccc}
\hline
spectral type &\multicolumn{4}{c}{dM1-2e $^1$}\\
$T_\mathrm{star}$&\multicolumn{4}{c}{3900K$^2$}\\
$i$&\multicolumn{4}{c}{$\approx70^\circ$$^2$}\\
$v\sin i$ & \multicolumn{4}{c}{61 km/s$^1$}\\
distance & \multicolumn{4}{c}{45.5 $\pm$2.1 pc$^2$}\\
$M_V$& \multicolumn{4}{c}{8.54 $\pm$ 0.12$^2$}\\
$r\sin i$ & \multicolumn{4}{c}{0.549 $\pm$ 0.002 R$_\odot$$^2$}\\
\hline
&$B$&$V$&$R_C$&$I_C$\\
\hline

unspotted $\Delta$ mag.&-0.470&-1.325&-1.930&-2.600\\
limb darkening$^3$&0.763&0.670&0.617&0.509\\
\hline
\label{tab:params}
\end{tabular}

\smallskip
$^1$:\cite{jeffries}\\
$^2$:\cite{barnes}\\
$^3$:\cite{vanhamme}
\end{table}
The light variation of EY Dra is caused by the rotation of the spotted surface.
The modelling of this spottedness was done using \sml \citep{sml}, which describes the intensity changes in an analytic way using homogeneous circular spots \citep{budding}. This assumption is a valid approximation e.g. to the shape of the sunspots or the stellar spots observed with Doppler-imaging. Another advantage of this model is the low number of free parameters: only the radius, location, and the temperature has to be fitted. 
"Spots" however are probably not homogeneous dark areas, but  mixtures of unresolved dark (cool) and bright (hot) regions, similar to the active nests observed on the Sun. Thus, the temperature of the model spots is regarded as the average temperature of hot and cool areas. 

During the modelling we have assumed two spots on the northern hemisphere, since   a two-spot model can follow arbitrary spotted light curves
well within the precision of the photometry \citep{spottest}. \sml is capable of spot temperature modelling making use of the colour index curves. We have chosen $V-I_C$ (Fig. \ref{fig:slices_ci}) for determining the spot temperatures, since this colour index variation relates to the temperature changes the best.
During the fitting process $T_\mathrm{eff}$ was set to 3900K according to \cite{barnes}.
Basic stellar parameters are summarized in Table \ref{tab:params}. The brightest values of the long-term light curve were used as unspotted magnitudes in each passband.
Modelling was run for 16 light curves, plotted in Fig. \ref{fig:slices}. 
Two samples of the phased $V$ light curve, $B-V$, $V-R_C$, and $V-I_C$ colour indices are plotted in Fig. \ref{fig:smlfit} together with the spot model fit. 
The fits show  model curves with parameters derived from the  modelling of the $V-I_C$ colour index curve, only the temperature-dependent parameters were changed for the plots.
The resulting parameters are plotted in the lower panels of Fig. \ref{fig:allv}, except spot latitudes, since these values cannot be precisely determined at the given photometric precision. The spot temperature is about 500K below the photospheric temperature. Due to the low amplitude of the light curves the derived temperatures have relatively large errors, therefore their change seen in Fig. \ref{fig:allv} has low significance. Models with different spot temperatures by 2-300K result in changing the spot radii by a few (1-3) degrees, so the overall picture of the spot radii variability remains the same.

The longitudes of both of the two active regions remain similar, i.e. stay within $\approx80^\circ$ during the time of the observations, more than 1000 days. 
 In this time, the active region at phase $\approx0.3$  slowly decays, while the other one, at phase $\approx0.8$ gets more and more prominent.

\subsection{Flares}

\begin{figure*}
\centering
\includegraphics[angle=0,width=0.4\textwidth]{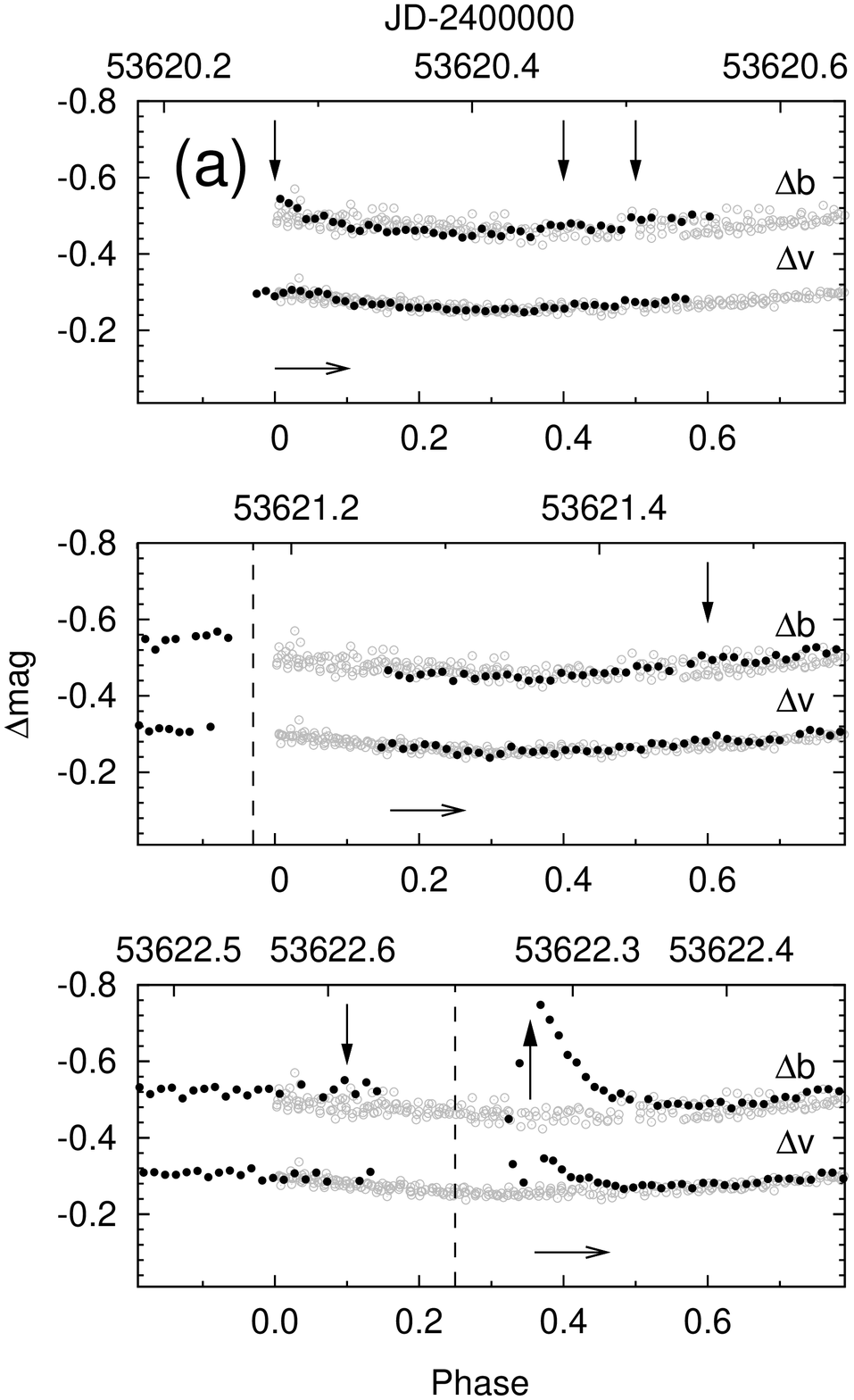}
\includegraphics[angle=0,width=0.4\textwidth]{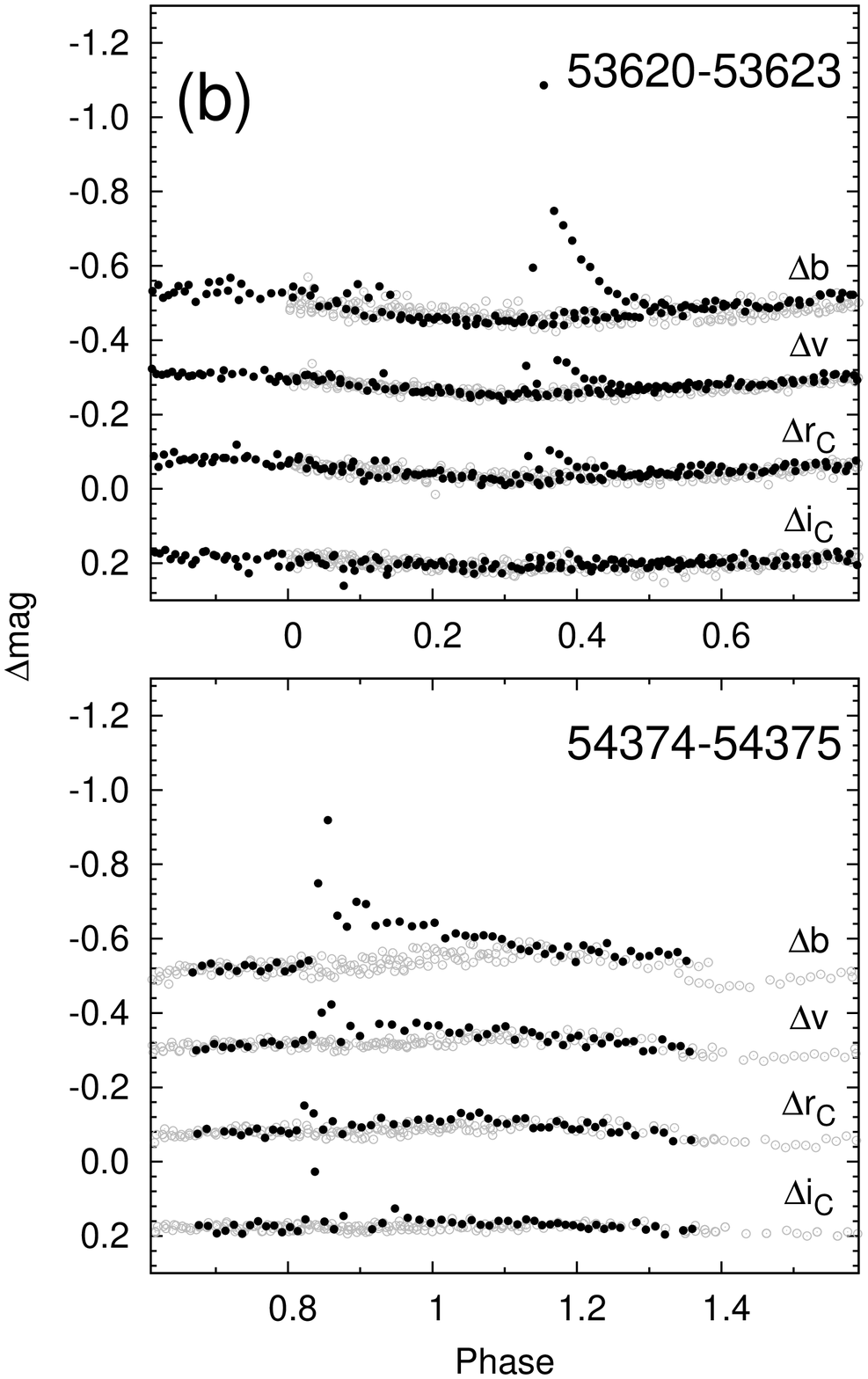}
\caption{
(a): Phased $bv$ light curves showing the three nights of the first flare event separately. Vertical arrows show the outbursts, dashed lines and the horizontal arrows at the bottom of the plots indicate the beginning of the observations.
(b): Phased $bv(ri)_C$ light curves showing two energetic flares on EY Dra. Gray points show additional observations of nearby nights  for comparison. $v$, $r_C$, and $i_C$ light curves are shifted arbitrarily.
Note, that the plots show instrumental values (see text).
}
\label{fig:flares}
\end{figure*}

\begin{table}
\caption{Estimated energies, equivalent durations of the two large flares, and quiescent fluxes in different filters.}
{
\centering
\begin{tabular}{cccccc}
\hline
&\multicolumn{2}{c}{Flare}&\multicolumn{2}{c}{Equivalent }&Quiescent\\
&\multicolumn{2}{c}{energy}&\multicolumn{2}{c}{duration}&flux\\
&\multicolumn{2}{c}{$[\times10^{34}$ ergs$]$}&\multicolumn{2}{c}{$[$sec$]$}&$[\times10^{31}$ ergs/sec$]$\\
Flare	& \#1	&	\#2	&\#1&\#2&\\
\hline
$B$   & 4.53 & 4.82 & 1232 & 1311 & 3.68\\
$V$   & 1.77 & 2.13 & 292  & 352  & 6.06\\
$R_C$ & 2.97 & 3.35 & 206  & 232  & 14.4\\
$I_C$ & 1.01 & 2.70 & 82   & 219  & 12.3\\
\hline
\end{tabular}
}
\smallskip\\
{\footnotesize
\#1: JD2453622, \#2: JD2454374
}
\label{tab:flare}
\end{table}

\begin{table}
\caption{Values for flare temperature $T_F$ and relative area $A = A_F /A_{\rm{star}}$ as a result of our trial-and-error procedure (see text).}
{
\centering
\begin{tabular}{l|llll}
\hline
Flare&    \multicolumn{2}{c}{\#1}       &  \multicolumn{2}{c}{\#2}\\
$T_{\rm flare}$  & $A_B$	 & $A_V$       & $A_B$     &   $A_V$\\
$[$K$]$ & \multicolumn{4}{c}{$[\%]$}\\
\hline
8000    &       1.04 & 0.29 & 0.56 & 0.13 \\
12000   &       0.25 & 0.09 & 0.13 & 0.04 \\
16000   &       0.12 & 0.04 & 0.06 & 0.02 \\
20000   &       0.07 & 0.03 & 0.04 & 0.01 \\
25000   &       0.05 & 0.02 & 0.03 & 0.01 \\
\hline
\end{tabular}
}
\smallskip\\
{ \footnotesize
\#1: JD2453622, \#2: JD2454374
}
\label{tab:flare_temp}
\end{table}

During the observations two noteworthy flare events happened: in JDs 2453622 and 2454374, lasting about 3.5--3.8 hours. Both events could be well seen in all filters, as plotted in Fig. \ref{fig:flares}. 
The plots show instrumental light curves, which was used for flare energy
calculations. The fast colour index changes does not allow a perfect
transformation to the international system during flares; the necessary
interpolations (which would be only rough approximations of the colour
indices anyway) would blur the details of the observations.
The first event is a complex one, covering three days. Before the large eruption five smaller outbursts were observed at different phases, probably not connected with the main event. All of them, however happened when the dominant active region was visible, around mid-phase of 0.3, suggesting a connection between the starspot and the flares.
According to its phase of about 0.9, (cf. Fig.~\ref{fig:slices} and \ref{fig:allv}), the other flare on JD 2454374 could also be associated with the smaller active region on EY Dra.

For estimating flare energies, we followed the method described by \cite{flare}. 
The energy emitted by the two flares and their equivalent durations are summarized in Tab. \ref{tab:flare}.
The second flare (JD 2454374) is slightly more energetic, though its peaks are lower in the red colours than that of the first flare (JD 2453622) (Fig. \ref{fig:flares}b).

We give a rough estimation for the colour temperature of the two flares at their maxima, by following the method of \cite{jager}.
The flare flux is significantly measurable in $B$, and barely in $V$, thus we use only those colours. If $A$ is the ratio between the projected area of the flaring region
and the visible stellar surface, then the flux ratio between the flaring and the quiescent star is
\begin{equation}
[F_{\rm S}(1-A)+F_{\rm F}A]/F_{\rm S}=1-A+AB_{\rm F}/B_{\rm S}
\label{eq:fluxrat}
\end{equation}
where $F_{\rm S}$ is the total flux of the quiescent star (Table \ref{tab:flare}), $F_{\rm F}$ is the flare flux, $B_{\rm F}$ and $B_{\rm S}$ are the blackbody fluxes of the flare and the star, respectively. Note that there are two unknown quantities in Eq. \ref{eq:fluxrat}, $A$ and $T_{\rm F}$, the flare temperature through $B_{\rm F}$. We only search for solutions by trial-and-error in a reasonable temperature range (see Table \ref{tab:flare_temp}). If we assume the given range feasible for a dMe star, we get some 0.1--0.01\% of the visible stellar surface for the flaring region, which is comparable with other results \citep[cf.][]{flare_area}.

\section{Discussion}
\begin{figure}
\centering
\includegraphics[angle=-90,width=.47\textwidth]{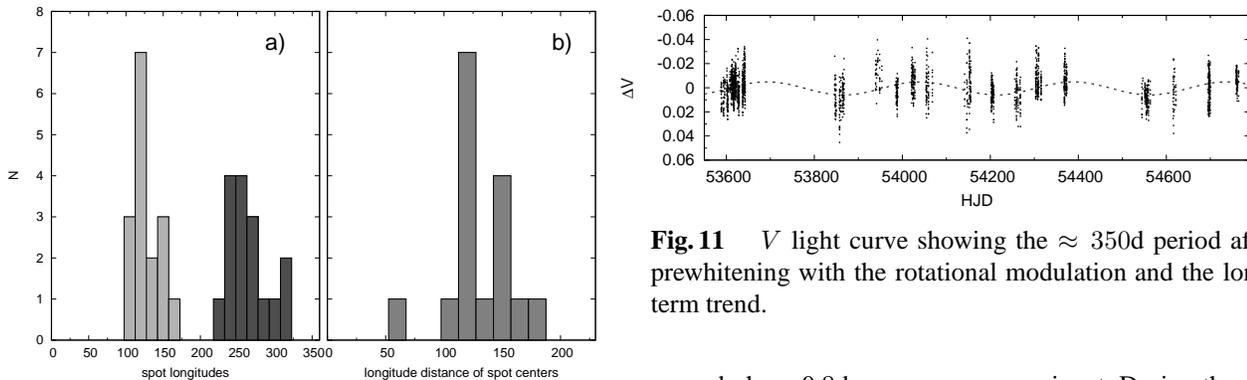}
\caption{Histograms showing the longitude distribution of the two active regions (a), and the longitude difference between the spot centers (b).}
\label{fig:histogram}
\end{figure}

\begin{figure}
\centering
\includegraphics[angle=-90,width=.48\textwidth]{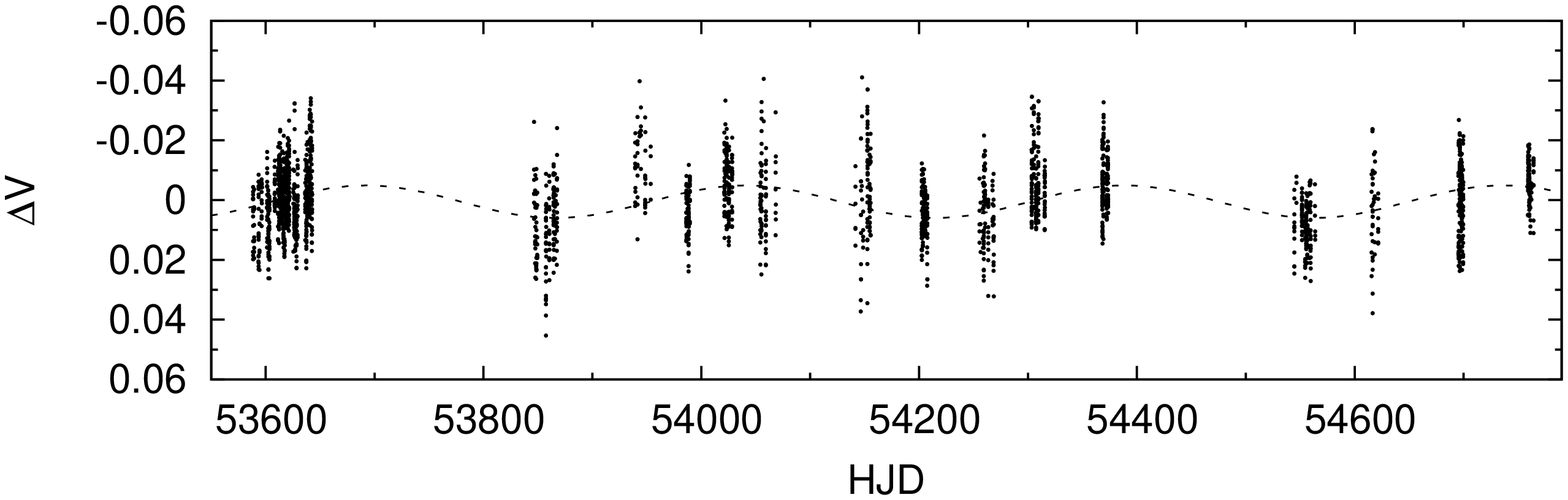}
\caption{$V$ light curve showing the $\approx350$d period after prewhitening with the rotational modulation and the long-term trend.}
\label{fig:long}
\end{figure}

Our spot modelling revealed two permanent, large active regions on EY Dra. 
From the beginning of the observations to JD $\approx$2454600 we can follow the decay of the spot group around phase 0.4 (see spot radii values on Fig. \ref{fig:allv} and phased light curves on Fig. \ref{fig:slices}). At the same time, the active region around phase 0.8 became more prominent. During the observations the total spotted area decreases from $\approx15\%$ to $\approx10\%$. Note, that the derived spottedness values are lower limits, since the unspotted brightness is set to the maximum observed magnitudes.
The change of the spot parameters in Fig. \ref{fig:allv} is reminiscent to those observed on FK\,Com by \cite{FKCom}.
The longitudes of the spots remain in a relatively narrow region, the distance of the two active regions is also similar, $\approx130^\circ$ (see the two histograms in Fig. \ref{fig:histogram}). 
The third panel of Fig. \ref{fig:allv} shows clearly, that after JD 2454400 the dominance between the two active region exchanges; the light curves of Fig.~\ref{fig:slices} directly show this change (JD 54363--54377 and JD 54539--54567).
This kind of so-called flip-flop phenomenon was observed on many stars, see e.g. \cite{flipflop0}. \cite{flipflop-bkg} showed that a mixed-mode dynamo including a non-axisymmetric and an oscillating axisymmetric mode can describe this kind of phenomenon in stars with thick convection zones.

The Fourier analysis showed, that there is one strong signal in the light curve. This modulation with $P_\mathrm{rot}=0.45875$d is caused by the rotation of the spotted surface.  
Since in most cases there are two active regions on EY Dra (see Fig. \ref{fig:slices}), a signal appears also at $P_\mathrm{rot}/2$ in the spectrum. Next to the strong signals of  $P_\mathrm{rot}$ and  $P_\mathrm{rot}/2$ weaker peaks can be found. This might be the result of differential rotation, but the difference between the signals is too small to draw such a conclusion. Differential rotation can still be present on the stellar surface, but when the latitudes of the active regions are close to each other, it is beyond the limit of detection. Unfortunately from photometry alone no trustworthy information can be determined on the spot latitudes. On the other hand, the small peaks near the rotational period and its half in the Fourier spectrum may originate simply from the modulation residuals, i.e., from the change of the light curve shapes. In summary: the detected rotational period of EY~Dra is remarkably stable during the three years of our observations. 
This could be the signature of a persisting magnetic configuration on the stellar surface, with spots on the same latitude. In the same time the two narrow regions of spot longitudes (see Fig. \ref{fig:histogram}a) also means long-term stability of the spot position. Stable magnetic fields lasting for at least one year was observed on V374 Peg \citep{morin}, which is about 0.3 solar mass and fully convective. EY Dra is also a low mass star  \citep[about 0.5 solar mass, cf.][]{eibe} with a very deep convection zone, therefore a similar stable magnetic configuration on its surface is possible.

Another peak in the amplitude spectrum indicates a long-term variation with $P_\mathrm{cycle}\approx350$d (see the light curve plotted in Fig. \ref{fig:long}). This can be a result of an activity cycle, similar to the 11-year long cycle observed on the Sun. Signs for the presence of this year-long cycle can also be found in ROTSE archive data, as shown in \cite{eydra}. This solar-like cycle is the shortest activity cycle known. As to our knowledge, there are only two M stars which show activity cycles: EY~Dra (M1-2) with a very fast rotation of 0.46 days and a cycle length of about 1 year, and HD~95735 (M2), slowly rotating with a period of about 55 days having cycles of 3.4--3.9 and $\approx$11 years (cf. \cite{cycles}. 

\section{Summary}
\begin{itemize}
\item Fourier analysis showed, that the rotation period is $P_\mathrm{rot}=0.45875$d, which is stable during the observations.
\item An activity cycle of $\approx350$d has been found.
\item Spot modelling showed that the active regions on the surface are located around $\approx130$ and $\approx250$ degrees.
\item The evolution of the surface indicates a possible flip-flop mechanism.
\item No reliable evidence was observed on differential rotation.
\item Two flare events were found around JDs 2453622 and on 2454374, possibly connected with the active regions on the surface.
\end{itemize}

\acknowledgements
This research has made use of the SIMBAD database, operated at CDS Strasbourg, France. The financial support of OTKA grant K-68626 and is acknowledged.
ZsK is a grantee of the Bolyai J\'anos Scholarship of the Hungarian Academy of Sciences.


\begin{thebibliography}{}
\bibitem[Ayres 
\& Linsky(1980)]{rot-act} Ayres, T.~R., \& Linsky, J.~L.\ 1980, \apj, 241, 279 
\bibitem[{Barnes \& Collier Cameron (2001)}]{barnes} Barnes J. R., Collier Cameron A., 2001, \mnras, 326, 950
\bibitem[Berdyugina(2005)]{review} Berdyugina, S.~V.\ 2005, 
Living Reviews in Solar Physics, 2, 8 
\bibitem[Budding(1977)]{budding} Budding, E.\ 1977, \apss, 48, 207 
\bibitem[Clarke(2002)]{sllk} Clarke, D.\ 2002, \aaa, 386, 763 
\bibitem[Eibe(1998)]{eibe}
Eibe M. T., 1998, \aaa, 337, 757

\bibitem[Elstner \& Korhonen(2005)]{flipflop-bkg} Elstner, D., \& Korhonen, H.\ 2005, Astronomische Nachrichten, 326, 278
\bibitem[Hall(2008)]{review-ca} Hall, J.~C.\ 2008, Living Reviews in Solar Physics, 5, 2 
\bibitem[de Jager et al.(1986)]{jager} de Jager, C., et al.\ 1986, \aaa, 156, 95
\bibitem[Jetsu et al.(1991)]{flipflop0} Jetsu, L., Pelt, J., 
Tuominen, I., 
\& Nations, H.\ 1991, IAU Colloq.~130: The Sun and Cool Stars.~Activity, Magnetism, Dynamos, 380, 381
\bibitem[Jeffries et al. (1994)]{jeffries}
Jeffries R. D., James D.J., Bromage G. E., 1994, \mnras, 271, 476
\bibitem[Koll\'ath (1990)]{mufran}
Koll\'ath, Z., 1990, The program package MUFRAN, Occasional Technical Notes of Konkoly Observatory, No. 1 \verb+www.konkoly.hu/Mitteilungen/+\\
\verb+Mitteilungen.html#TechNotes+
\bibitem[Korhonen et al.(2007)]{korhonen} Korhonen, H., Brogaard, K., Holhjem, K., Ramstedt, S., Rantala, J., Th{\"o}ne, C.~C., \& Vida, K.\ 2007, Astronomische Nachrichten, 328, 897 
\bibitem[K\H{o}v\'ari \& Bartus(1997)]{spottest} K\H{o}v\'ari, Zs., \& Bartus, J.\ 1997, \aaa, 323, 801 
\bibitem[K\H{o}v{\'a}ri et al.(2007)]{flare} K\H{o}v{\'a}ri, Zs., 
Vilardell, F., Ribas, I., Vida, K., van Driel-Gesztelyi, L., Jordi, C., 
\& Ol{\'a}h, K.\ 2007, Astronomische Nachrichten, 328, 904 
\bibitem[Morin et al.(2008)]{morin} Morin, J., et al.\ 2008, \mnras, 384, 77 
\bibitem[Ol{\'a}h et al.(2006)]{cycles} Ol{\'a}h, K., Koll{\'a}th, Z., Granzer, T., Strassmeier, K. G., Lanza, A. F., Järvinen, S., Korhonen, H., Baliunas, S. L., Soon, W., Messina, S., Cutispoto, G.\ 2009, \aaa, 501, 703 
\bibitem[Ol{\'a}h et al.(2006)]{FKCom} Ol{\'a}h, K., Korhonen, H., K{\H o}v{\'a}ri, Z., Forg{\'a}cs-Dajka, E., \& Strassmeier, K.~G.\ 2006, \aaa, 452, 303
\bibitem[Ol{\'a}h et al.(2001)]{flare_area} Ol{\'a}h, K., Strassmeier, K.~G., Kov{\'a}ri, Z., \& Guinan, E.~F.\ 2001, \aaa, 372, 119 
\bibitem[Pagano et 
al.(2001)]{euv} Pagano, I., Rodon{\`o}, M., Linsky, J.~L., Neff, J.~E., Walter, F.~M., Kov{\'a}ri, Z., \& Matthews, L.~D.\ 2001, \aaa, 365, 128 
\bibitem[Pizzolato et 
al.(2003)]{rot-mag} Pizzolato, N., Maggio, A., Micela, G., Sciortino, S., \& Ventura, P.\ 2003, \aaa, 397, 147 
\bibitem[Plavchan et al.(2009)]{spitzer} Plavchan, P., Werner, 
M.~W., Chen, C.~H., Stapelfeldt, K.~R., Su, K.~Y.~L., Stauffer, J.~R., 
\& Song, I.\ 2009, \apj, 698, 1068 
\bibitem[Press et al.(1992)]{error} Press, W. H., Teukolsky, S. A., Vetterling, W. T., \& Flannery, B. P. 1992,
Numerical Recepies in FORTRAN: The Art of Scientific Computing, 2nd ed. (New York: Cambridge Univ. press)
\bibitem[Rib{\'a}rik et al.(2003)]{sml} Rib{\'a}rik, G., Ol{\'a}h, K., \& Strassmeier, K.~G.\ 2003, Astronomische Nachrichten, 324, 202 
\bibitem[Robb \& Cardinal(1995)]{robb} Robb, R.~M., \& Cardinal, R.~D.\ 1995, Information Bulletin on Variable Stars, 4270, 1
\bibitem[Sch\"ussler \& Solanki(1992)]{polar_cap} Sch\"ussler, M., \& Solanki, S.~K.\ 1992, \aaa, 264, L13
\bibitem[van Hamme(1993)]{vanhamme} van Hamme, W.\ 1993, \aj, 106, 2096 
\bibitem[Vida(2007)]{eydra} Vida, K.\ 2007, Astronomische 
Nachrichten, 328, 817 

\end{thebibliography}
\end{document}